# Proof of non-extremal nature of excited-state energies in nonrelativistic quantum mechanics with the use of constrained derivatives

Tamás Gál

Quantum Theory Project, Department of Physics, University of Florida,
Gainesville, Florida 32611, USA

**Abstract:** With the use of a second derivative test based on constrained second derivatives, a proof is given that excited states in nonrelativistic quantum mechanics are saddle points of the energy expectation value, and is shown, further, how to determine their Morse index.

*E-mail: gal@qtp.ufl.edu

It is one of the most well-known, basic facts of nonrelativistic quantum mechanics that the ground state always minimizes the energy expectation value $\langle\psi|\hat{H}|\psi\rangle$ under the normalization constraint

$$\langle\psi|\psi\rangle=1 . \tag{1}$$

This is the basis for the Rayleigh-Ritz variational method, and has been the most widely applied tool for electronic structure calculations since the birth of quantum mechanics. As regards excited states, it is also well-known [1,2] that (i) all eigenstates of a given Hamiltonian $\hat{H}$ emerge as stationary points of $\langle\psi|\hat{H}|\psi\rangle$ subject to Eq.(1), and (ii) the $n$th excited state of $\hat{H}$ minimizes $\langle\psi|\hat{H}|\psi\rangle$ subject to the constraint that $|\psi\rangle$ is orthogonal to the first $n$-1 eigenstates of $\hat{H}$ in addition to Eq.(1). However, it is less known fact what is the nature of the stationary points of $\langle\psi|\hat{H}|\psi\rangle$ without the orthogonality constraints – whether at the eigenstates of $\hat{H}$, $\langle\psi|\hat{H}|\psi\rangle$ has local minima, or there are not local extrema but $\langle\psi|\hat{H}|\psi\rangle$ has a metastable character at the excited states. In this paper, we will present a proof of the latter with the use of a second derivative test based on constrained differentiation [3], and further, show that the index of the saddle point a given excited state is equals the total number of (linearly independent) lower-lying energy eigenstates.

The concept of constrained derivatives [3,4]

$$\frac{\delta A[\rho]}{\delta_C \rho(x)} = \frac{\delta A[\rho]}{\delta \rho(x)} - \mu[C[\rho]; A[\rho]]\frac{\delta C[\rho]}{\delta \rho(x)} \tag{2}$$

can be considered as a generalization of the method of Lagrange multipliers [5] for non-stationary situations in the presence of constraints $C[\rho]=0$, in the sense that Lagrange's method leads to a modification

$$\frac{\delta A[\rho]}{\delta \rho(x)} - \mu\frac{\delta C[\rho]}{\delta \rho(x)} = 0 \tag{3}$$

of the Euler equation $\frac{\delta A[\rho]}{\delta \rho(x)}=0$ determining the stationary points of a functional $A[\rho]$. On the other hand, the philosophies behind Eqs.(2) and (3) are different, since (i) Eq.(3) is obtained from a modification of the functional $A[\rho]$, $A[\rho]-\mu C[\rho]$, while constrained derivatives modify the differentiation itself, and (ii) the Lagrange multiplier $\mu$ of Eq.(3) is undetermined in the sense that its value is determined only after the solution of Eq.(3) for $\rho(x)$, by adjusting $\mu$ for $\rho(x)$ to satisfy the constraint, while the multiplier $\mu$ in Eq.(2) is explicitly determined by the given functional and the constraint. Constrained derivatives can be useful in the construction of dynamical models where the evolution of variables is restricted by conservation constraints, as in the binary thin-film model of Clarke [6], experimentally verified by Thomas et al. [7], and in the stability analysis of equilibrium states under constraints, as in the constrained density functional theory study of droplet and bubble nucleation and growth by Uline et al. [8,9]. The appearance of constrained derivatives in dynamical equations can be explained by an invariance principle regarding the form of the equations [10], while the role of constrained (second) derivatives in stability analysis was established in [11]. As observed in [9], constrained derivatives have (numerical) use even in the determination of stationary points, through

$$\frac{\delta A[\rho]}{\delta_C \rho(x)} = 0 \ , \tag{4}$$

where otherwise their application is mathematically equivalent with the use of the usual method of Lagrange multipliers [3,4].

As has been shown in [11], under constraints, in the place of the well-known necessary condition

$$\iint \frac{\delta^2 A[\rho]}{\delta \rho(x) \delta \rho(x')} \Delta\rho(x) \Delta\rho(x') dx dx' \qquad \text{be non-negative} \tag{5a}$$

and sufficient condition

$$\iint \frac{\delta^2 A[\rho]}{\delta \rho(x) \delta \rho(x')} \Delta\rho(x) \Delta\rho(x') dx dx' \qquad \text{be strongly positive} \tag{5b}$$

for having a local minimum of $A[\rho]$ [5], we have

$$\iint \frac{\delta^2 A[\rho]}{\delta_C \rho(x) \delta_C \rho(x')} \Delta\rho(x) \Delta\rho(x') dx dx' \qquad \text{be non-negative} \tag{6a}$$

and

$$\iint \frac{\delta^2 A[\rho]}{\delta_C \rho(x) \delta_C \rho(x')} \Delta\rho(x) \Delta\rho(x') dx dx' \qquad \text{be strongly positive} \ , \tag{6b}$$

respectively. Eq.(6b) leads to the practically implementable sufficient condition that the eigenvalues of

$$\int \frac{\delta^2 A[\rho]}{\delta_C \rho(x) \delta_C \rho(x')} \Delta\rho(x') dx' = \lambda \Delta\rho(x) \tag{7}$$

be strongly positive, $\lambda \geq p > 0$, for a local minimum, while from Eq.(6a), it follows that if there are both positive and negative eigenvalues among the $\lambda$'s, we have a saddle point of $A[\rho]$ at the given $\rho(x)$. This condition will be applied in the following to determine whether the stationary points of $\langle\psi|\hat{H}|\psi\rangle$ subject to Eq.(1), i.e. the eigenstates of the Schrödinger equation, are local extrema or saddle points of $\langle\psi|\hat{H}|\psi\rangle$ in the space of normalized wavefunctions.

On the basis of the expansion

$$A[\rho+\Delta_C\rho]=A[\rho]+\int\frac{\delta A[\rho]}{\delta_C\rho(x)}\Delta\rho(x)dx+\frac{1}{2!}\iint\frac{\delta^2 A[\rho]}{\delta_C\rho(x)\delta_C\rho(x')}\Delta\rho(x)\Delta\rho(x')dxdx'+...\,, \qquad (8)$$

derived in [11] for functionals over a general Banach space of (one- or multi-valued) functions $\rho(x)$ in the presence of a constraint $C$, the increment of a real-valued functional $F[\psi]=A[\psi,\psi^*]$ of a complex variable $\psi(x)$ due to a normalization-conserving change of $\psi(x)$ can be given as

$$A[\psi+\Delta_n\psi,\psi^*+\Delta_n\psi^*]-A[\psi,\psi^*]=\int\frac{\delta A[\psi,\psi^*]}{\delta_n\psi(x)}\Delta\psi(x)dx+\int\frac{\delta A[\psi,\psi^*]}{\delta_n\psi^*(x)}\Delta\psi^*(x)dx$$

$$+\frac{1}{2}\iint\frac{\delta^2 A[\psi,\psi^*]}{\delta_n\psi(x)\delta_n\psi(x')}\Delta\psi(x)\Delta\psi(x')dxdx'+\iint\frac{\delta^2 A[\psi,\psi^*]}{\delta_n\psi(x)\delta_n\psi^*(x')}\Delta\psi(x)\Delta\psi^*(x')dxdx'$$

$$+\frac{1}{2}\iint\frac{\delta^2 A[\psi,\psi^*]}{\delta_n\psi^*(x)\delta_n\psi^*(x')}\Delta\psi^*(x)\Delta\psi^*(x')dxdx'+... \qquad (9)$$

($x$ embraces all the variables of $\psi$.) The constrained first and second derivatives in Eq.(9) can be obtained [3,4,11] as the unconstrained first and second derivatives of

$$A[\psi_n[\psi,\psi^*],\psi_n^*[\psi,\psi^*]] \qquad (10)$$

(taken at $\int\psi^*(x)\psi(x)dx=1$), respectively, where $\psi_n[\psi,\psi^*]$ is a functional that gives a value $\psi_n(x)$ that (i) satisfies Eq.(1), for any input $\psi(x)$, and (ii) is identical with the input if that already fulfils Eq.(1), that is, $\psi_n[\tilde\psi_n,\tilde\psi_n^*]=\tilde\psi_n$. The simplest choice of $\psi_n[\psi,\psi^*]$ is

$$\psi_n[\psi,\psi^*]=\frac{\psi(x)}{\sqrt{\int\psi^*(x')\psi(x')dx'}}\,, \qquad (11)$$

which fulfils the degree-zero homogeneity condition [4] as well, in addition to conditions (i) and (ii) above; i.e., $\psi_n[k\psi,k\psi^*]=\psi_n[\psi,\psi^*]$ for any real number $k$.

For generality, we will determine the constrained derivatives for a general two-variable functional $A[a,b]$ with the constraint

$$\int b(x)a(x)dx=n\,, \qquad (12)$$

obtaining the needed constrained derivatives by the choice of $a(x) := \psi(x)$ and $b(x) := \psi^*(x)$, with $n=1$. For this, the functional

$$A[a_n[a,b], b_n[a,b]] \tag{13}$$

needs to be differentiated, where

$$a_n[a,b] = \sqrt{\frac{n}{\int a(x)b(x)dx}} \, a(x) \tag{14a}$$

and

$$b_n[a,b] = \sqrt{\frac{n}{\int a(x)b(x)dx}} \, b(x) \; . \tag{14b}$$

The first derivatives of Eq.(13) emerge as

$$\frac{\delta A[a_n[a,b], b_n[a,b]]}{\delta a(x)} = \int \frac{\delta A}{\delta a_n(x')} \frac{\delta a_n(x')}{\delta a(x)} dx' + \int \frac{\delta A}{\delta b_n(x')} \frac{\delta b_n(x')}{\delta a(x)} dx' \; , \tag{15a}$$

$$\frac{\delta A[a_n[a,b], b_n[a,b]]}{\delta b(x)} = \int \frac{\delta A}{\delta a_n(x')} \frac{\delta a_n(x')}{\delta b(x)} dx' + \int \frac{\delta A}{\delta b_n(x')} \frac{\delta b_n(x')}{\delta b(x)} dx' \; , \tag{15b}$$

with

$$\frac{\delta a_n(x')[a,b]}{\delta a(x)} = \sqrt{\frac{n}{\int a(x)b(x)dx}} \left( \delta(x'-x) - \frac{a(x')b(x)}{2\int a(x)b(x)dx} \right) , \tag{16a}$$

$$\frac{\delta a_n(x')[a,b]}{\delta b(x)} = -\sqrt{\frac{n}{\int a(x)b(x)dx}} \frac{a(x')a(x)}{2\int a(x)b(x)dx} \; , \tag{16b}$$

$$\frac{\delta b_n(x')[a,b]}{\delta a(x)} = -\sqrt{\frac{n}{\int a(x)b(x)dx}} \frac{b(x')b(x)}{2\int a(x)b(x)dx} \; , \tag{16c}$$

$$\frac{\delta b_n(x')[a,b]}{\delta b(x)} = \sqrt{\frac{n}{\int a(x)b(x)dx}} \left( \delta(x'-x) - \frac{b(x')a(x)}{2\int a(x)b(x)dx} \right) . \tag{16d}$$

Taking Eqs.(15) at $(a(x), b(x))$ that satisfy the constraint Eq.(12) (i.e., formally, substituting $n = \int b(x)a(x)dx$ in Eqs.(15)), we obtain

$$\frac{\delta A[a,b]}{\delta_n a(x)} = \frac{\delta A[a,b]}{\delta a(x)} - \frac{b(x)}{2n} \left( \int a(x') \frac{\delta A[a,b]}{\delta a(x')} dx' + \int b(x') \frac{\delta A[a,b]}{\delta b(x')} dx' \right) \tag{17a}$$

and

$$\frac{\delta A[a,b]}{\delta_n b(x)} = \frac{\delta A[a,b]}{\delta b(x)} - \frac{a(x)}{2n} \left( \int a(x') \frac{\delta A[a,b]}{\delta a(x')} dx' + \int b(x') \frac{\delta A[a,b]}{\delta b(x')} dx' \right) \tag{17b}$$

as the constrained first derivatives. Differentiating Eqs.(15), with Eqs.(16), with respect to $a(x)$ and $b(x)$, then substituting $n = \int b(x) a(x) dx$, the constrained second derivatives can be obtained,

$$\frac{\delta^2 A[a,b]}{\delta_n a(x) \delta_n a(x')}$$

$$= \frac{\delta^2 A[a,b]}{\delta a(x) \delta a(x')} - \frac{b(x)}{2} \int a(x'') \frac{\delta^2 A[a,b]}{\delta a(x'') \delta a(x')} dx'' - \frac{b(x')}{2} \int a(x'') \frac{\delta^2 A[a,b]}{\delta a(x'') \delta a(x)} dx'' - b(x') \int b(x'') \frac{\delta^2 A[a,b]}{\delta a(x) \delta b(x'')} dx''$$

$$+ \frac{b(x)b(x')}{4} \left( \iint a(x'')a(x''') \frac{\delta^2 A[a,b]}{\delta a(x'') \delta a(x''')} dx'' dx''' + 2 \iint a(x'')b(x''') \frac{\delta^2 A[a,b]}{\delta a(x'') \delta b(x''')} dx'' dx''' + \iint b(x'')b(x''') \frac{\delta^2 A[a,b]}{\delta b(x'') \delta b(x''')} dx'' dx''' \right)$$

$$- \frac{b(x)}{2} \frac{\delta A[a,b]}{\delta a(x')} - \frac{b(x')}{2} \frac{\delta A[a,b]}{\delta a(x)} + \frac{3}{4} b(x)b(x') \left( \int a(x'') \frac{\delta A[a,b]}{\delta a(x'')} dx'' + \int b(x'') \frac{\delta A[a,b]}{\delta b(x'')} dx'' \right) , \quad (18a)$$

$$\frac{\delta^2 A[a,b]}{\delta_n a(x) \delta_n b(x')}$$

$$= \frac{\delta^2 A[a,b]}{\delta a(x) \delta b(x')} - \frac{b(x)}{2} \left( \int a(x'') \frac{\delta^2 A[a,b]}{\delta a(x'') \delta b(x')} dx'' + \int b(x'') \frac{\delta^2 A[a,b]}{\delta b(x'') \delta b(x')} dx'' \right)$$

$$- \frac{a(x')}{2} \left( \int a(x'') \frac{\delta^2 A[a,b]}{\delta a(x'') \delta a(x)} dx'' + \int b(x'') \frac{\delta^2 A[a,b]}{\delta b(x'') \delta a(x)} dx'' \right)$$

$$+ \frac{b(x)a(x')}{4} \left( \iint a(x'')a(x''') \frac{\delta^2 A[a,b]}{\delta a(x'') \delta a(x''')} dx'' dx''' + 2 \iint a(x'')b(x''') \frac{\delta^2 A[a,b]}{\delta a(x'') \delta b(x''')} dx'' dx''' + \iint b(x'')b(x''') \frac{\delta^2 A[a,b]}{\delta b(x'') \delta b(x''')} dx'' dx''' \right)$$

$$- \frac{a(x')}{2} \frac{\delta A[a,b]}{\delta a(x)} - \frac{b(x)}{2} \frac{\delta A[a,b]}{\delta b(x')} + \left( \frac{3}{4} b(x)a(x') - \frac{\delta(x-x')}{2} \right) \left( \int a(x'') \frac{\delta A[a,b]}{\delta a(x'')} dx'' + \int b(x'') \frac{\delta A[a,b]}{\delta b(x'')} dx'' \right) , \quad (18b)$$

and

$$\frac{\delta^2 A[a,b]}{\delta_n b(x) \delta_n b(x')}$$

$$= \frac{\delta^2 A[a,b]}{\delta b(x) \delta b(x')} - \frac{a(x)}{2} \int b(x'') \frac{\delta^2 A[a,b]}{\delta b(x'') \delta b(x')} dx'' - \frac{a(x')}{2} \int b(x'') \frac{\delta^2 A[a,b]}{\delta b(x'') \delta b(x)} dx'' - a(x') \int a(x'') \frac{\delta^2 A[a,b]}{\delta b(x) \delta a(x'')} dx''$$

$$+ \frac{a(x)a(x')}{4} \left( \iint a(x'')a(x''') \frac{\delta^2 A[a,b]}{\delta a(x'') \delta a(x''')} dx'' dx''' + 2 \iint a(x'')b(x''') \frac{\delta^2 A[a,b]}{\delta a(x'') \delta b(x''')} dx'' dx''' + \iint b(x'')b(x''') \frac{\delta^2 A[a,b]}{\delta b(x'') \delta b(x''')} dx'' dx''' \right)$$

$$- \frac{a(x)}{2} \frac{\delta A[a,b]}{\delta b(x')} - \frac{a(x')}{2} \frac{\delta A[a,b]}{\delta b(x)} + \frac{3}{4} a(x)a(x') \left( \int a(x'') \frac{\delta A[a,b]}{\delta a(x'')} dx'' + \int b(x'') \frac{\delta A[a,b]}{\delta b(x'')} dx'' \right) . \quad (18c)$$

To simplify presentation, in Eqs.(18), $n$ of Eq.(12) is already taken to be 1.

Applying the above formulae for a functional of the form $F[\psi] = A[\psi, \psi^*] = \int \psi^*(x) \hat{A} \psi(x) dx$, where $\hat{A}$ Hermitian, and utilizing its linearity both in $\psi(x)$ and in $\psi^*(x)$, we obtain

$$\int \frac{\delta^2 A[\psi,\psi^*]}{\delta_n\psi^*(x)\delta_n\psi(x')}\Delta\psi(x')dx' = \hat{A}\Delta\psi(x) - \frac{1}{2}\psi(x)\int \psi^*(x')\hat{A}\Delta\psi(x')dx' - \frac{1}{2}\int \psi^*(x')\Delta\psi(x')dx'\,\hat{A}\psi(x)$$
$$+\frac{1}{2}\psi(x)\int \psi^*(x')\Delta\psi(x')dx'\int \psi^*(x')\hat{A}\psi(x')dx' - \frac{1}{2}\int \psi^*(x')\Delta\psi(x')dx'\,\hat{A}\psi(x) - \frac{1}{2}\psi(x)\int \psi^*(x')\hat{A}\Delta\psi(x')dx'$$
$$+\left(\frac{3}{2}\psi(x)\int \psi^*(x')\Delta\psi(x')dx' - \Delta\psi(x)\right)\int \psi^*(x')\hat{A}\psi(x')dx' \tag{19a}$$

and

$$\int \frac{\delta^2 A[\psi,\psi^*]}{\delta_n\psi^*(x)\delta_n\psi^*(x')}\Delta\psi^*(x')dx' = -\int \psi(x')\Delta\psi^*(x')dx'\,\hat{A}\psi(x) + \frac{1}{2}\psi(x)\int \psi(x')\Delta\psi^*(x')dx'\int \psi^*(x')\hat{A}\psi(x')dx'$$
$$+\frac{3}{2}\psi(x)\int \Delta\psi^*(x')\psi(x')dx'\int \psi^*(x')\hat{A}\psi(x')dx' - \frac{1}{2}\int \Delta\psi^*(x')\psi(x')dx'\,\hat{A}\psi(x) - \frac{1}{2}\psi(x)\int \Delta\psi^*(x')\hat{A}\psi(x')dx' \;.$$
(19b)

$\int \frac{\delta^2 A[\psi,\psi^*]}{\delta_n\psi(x)\delta_n\psi^*(x')}\Delta\psi^*(x')dx'$ and $\int \frac{\delta^2 A[\psi,\psi^*]}{\delta_n\psi(x)\delta_n\psi(x')}\Delta\psi(x')dx'$ are the complex conjugates of

Eq.(19a) and Eq.(19b), respectively. Now, utilizing that $\psi(x)$ is the $k$th eigenstate of $\hat{A}$, i.e. $\hat{A}\psi(x) = A_k\psi(x)$,

$$\int \frac{\delta^2 A[\psi,\psi^*]}{\delta_n\psi^*(x)\delta_n\psi(x')}\Delta\psi(x')dx' = \hat{A}\Delta\psi(x) - A_k\Delta\psi(x) + \frac{1}{2}\psi(x)\left(A_k\int \psi^*(x')\Delta\psi(x')dx' - \int \psi^*(x')\hat{A}\Delta\psi(x')dx'\right)$$
(20a)

and

$$\int \frac{\delta^2 A[\psi,\psi^*]}{\delta_n\psi^*(x)\delta_n\psi^*(x')}\Delta\psi^*(x')dx' = 0 \tag{20b}$$

emerge. With this, we can write Eq.(9) as

$$A[\psi+\Delta_n\psi,\psi^*+\Delta_n\psi^*] - A[\psi,\psi^*] = \iint \frac{\delta^2 A[\psi,\psi^*]}{\delta_n\psi(x)\delta_n\psi^*(x')}\Delta\psi(x)\Delta\psi^*(x')dxdx' + \ldots, \tag{21}$$

where Eq.(4), i.e.

$$\frac{\delta A[\psi,\psi^*]}{\delta_n\psi(x)} = 0 \qquad \text{and} \qquad \frac{\delta A[\psi,\psi^*]}{\delta_n\psi^*(x)} = 0 \;, \tag{22}$$

following from the fact that the examined $\psi(x)$'s are the stationary points of $A[\psi,\psi^*] = \int \psi^*(x)\hat{A}\psi(x)dx$ subject to Eq.(1), has also been utilized. The eigenvalue equation corresponding to Eq.(7) will then be

$$\int \frac{\delta^2 A[\psi,\psi^*]}{\delta_n\psi^*(x)\delta_n\psi(x')}\Delta\psi(x')dx' = \lambda\,\Delta\psi(x) \;. \tag{23}$$

In the case of the Hamilton operator, with the property $\int \phi(x)\hat{H}\psi(x)\,dx = \int \hat{H}\phi(x)\psi(x)\,dx$, the two terms between the brackets in Eq.(20a) cancel each other. Thus, Eq.(23) gives

$$\hat{H}\Delta\psi(x) - E_k \Delta\psi(x) = \lambda \Delta\psi(x) \ . \tag{24}$$

The solutions $(\Delta\psi(x))_m$ of this eigenvalue equation are simply the eigenfunctions of the given $\hat{H}$, with eigenvalues

$$\lambda_m = E_m - E_k \ . \tag{25}$$

From Eq.(25), then, it can be seen that apart from the ground state, all eigenstates of $\hat{H}$ will only be saddle points of $\int \psi^*(x)\hat{H}\psi(x)\,dx$ (subject to Eq.(1)), and not local minimum points, since

$$\lambda_m < 0 \quad \text{for} \quad m=0,1,\ldots,k\text{-}1 \ , \tag{26a}$$

while

$$\lambda_m > 0 \quad \text{for} \quad m > k \ . \tag{26b}$$

As regards other operators than $\hat{H}$, the bracketed two terms in Eq.(20a) still vanish together, since $\Delta\psi(x)$ can be expanded into the eigenfunctions of the given $\hat{A}$, $\Delta\psi(x) = \sqrt{\kappa}\sum_j c_j \psi_j(x)$ (where the real multiplier $\sqrt{\kappa}$ appears due to the fact that $\Delta\psi(x)$ is not normalized), with the use of which $\int \psi^*(x')\hat{A}\Delta\psi(x')\,dx' = A_k c_k \sqrt{\kappa}$, cancelling the other term. Thus, the above result can be generalized for any $\int \psi^*(x)\hat{A}\psi(x)\,dx$ with Hermitian $\hat{A}$. We note here that, of course, the insertion of Eq.(11) into $\int \psi^*(x)\hat{H}\psi(x)\,dx$ yields the usual expression used to ensure the normalization of the wavefunction in the energy minimization; so one might raise that it is natural that the second derivative of this expression provides the proper substitute of the unconstrained second derivative in the presence of the constraint Eq.(1). This is true; however, proving the theorems Eq.(6) [11] would still be necessary – either one considers the derivatives gained via Eq.(11) constrained derivatives or not. (The idea behind constrained differentiation, practically, is the recognition that $\dfrac{\int \psi^*(x)\hat{H}\psi(x)\,dx}{\int \psi^*(x)\psi(x)\,dx}$ can be considered as a modification Eq.(11) of the functional variable of $\int \psi^*(x)\hat{H}\psi(x)\,dx$, making it possible to account for constraints in this way in the case of functionals of a general form.)

By recognizing an important property of the *n*-conserving derivatives Eq.(17), it is possible to determine on the basis of Eq.(26) what type of saddle point a *k*th eigenfunction of $\hat{A}$ is. In the Morse theory of (hyper)surfaces [12], an index is associated with every nondegenerate saddle point according to the number of negative eigenvalues of the Hessian corresponding to the considered stationary point of a given real-valued functional $F[\rho]$. The index of a stationary point gives the number of independent directions along which $F[\rho]$ decreases. In the case of a nondegenerate stationary point, a zero index signifies a local minimum, while positive indices (of less than the total number of the eigenvalues of the Hessian) correspond to saddles of different shapes. Without a constraint, this theory applies for the eigenvalues of $\int dx' \frac{\delta^2 F[\rho]}{\delta\rho(x)\delta\rho(x')}$; however, it is a question whether the eigenvalues of Eq.(7) are the proper choice to gain a Morse index for the constrained case, considering especially that there is an ambiguity in defining the constrained second derivative entering Eq.(7) [11], implying a possible variety of different choices to calculate an index.

As can be checked easily, Eq.(17) has the following property:

$$\int a(x)\frac{\delta A[a,b]}{\delta_n a(x)}dx + \int b(x)\frac{\delta A[a,b]}{\delta_n b(x)}dx = 0 , \qquad (27)$$

which gives

$$\int \psi(x)\frac{\delta A[\psi,\psi^*]}{\delta_n \psi(x)}dx + \int \psi^*(x)\frac{\delta A[\psi,\psi^*]}{\delta_n \psi^*(x)}dx = 0 \qquad (28)$$

for the case of interest. Applying Eq.(28) for the first derivative $\frac{\delta A[\psi,\psi^*]}{\delta_n \psi(x')}$ (in the place of $A[\psi,\psi^*]$), we have

$$\int \psi(x)\frac{\delta^2 A[\psi,\psi^*]}{\delta_n\psi(x)\delta_n\psi(x')}dx + \int \psi^*(x)\frac{\delta^2 A[\psi,\psi^*]}{\delta_n\psi^*(x)\delta_n\psi(x')}dx = 0 . \qquad (29)$$

Now, utilizing this result and (the complex conjugate of) Eq.(20b) for Eq.(23) multiplied by $\psi^*(x)$ and integrated over $x$, we obtain

$$\lambda_m \int \psi^*(x)(\Delta\psi(x))_m dx = 0 \qquad (30)$$

for the eigenfunctions and the corresponding eigenvalues of Eq.(23). [That is, the eigenfunctions of Eq.(23) should be either orthogonal to the given $\psi(x)$ or must correspond to $\lambda_m = 0$. Since the considered $\psi(x)$ are the eigenfunctions of $\hat{A}$ and the eigenfunctions of Eq.(23) are just the eigenfunctions of $\hat{A}$ (times some real constant), too, with eigenvalues Eq.(25), this is evidently true.] From Eq.(30), it follows that all eigenfunctions of Eq.(23) corresponding to a nonzero eigenvalue satisfy

$$\int \psi^*(x)\Delta\psi(x)dx + \int \psi(x)\Delta\psi^*(x)dx = 0 \ . \tag{31}$$

Eq.(31) can be obtained from

$$\int \psi^*(x)\Delta\psi(x)dx + \int \psi(x)\Delta\psi^*(x)dx = -\int \Delta\psi(x)\Delta\psi^*(x)dx \ , \tag{32}$$

(arising from Eq.(1) for an *n*-conserving change of $\psi(x)$) by neglecting the second-order term, and is valid for first-order *n*-conserving changes of $\psi(x)$.

A general increment $\Delta\psi(x)$ can be written with the help of orthogonal eigenfunctions of Eq.(23) as

$$\Delta\psi(x) = \sum_{i,j} c_{ij}(\Delta\psi(x))_{ij} \ , \tag{33}$$

where a second index has been introduced for possible degeneracy of a given eigenvalue $\lambda_i$. With the use of this,

$$\int \frac{\delta^2 A[\psi,\psi^*]}{\delta_n\psi^*(x)\delta_n\psi(x')}\Delta\psi(x')dx' = \sum_{i,j} \lambda_i c_{ij}(\Delta\psi(x))_{ij} \ . \tag{34}$$

Inserting Eq.(34) into Eq.(21), then, the orthogonality of the eigenfunctions,

$$\int (\Delta\psi^*(x))_{ij}(\Delta\psi(x))_{kl} dx = 0 \qquad (ij \neq kl) \ , \tag{35}$$

can be utilized to obtain

$$A[\psi+\Delta_n\psi, \psi^*+\Delta_n\psi^*] - A[\psi,\psi^*] = \sum_{i,j} \lambda_i |c_{ij}|^2 \int |(\Delta\psi(x))_{ij}|^2 dx + \ldots \ . \tag{36}$$

Since any eigenfunction $(\Delta\psi(x))_{ij}$ that does not satisfy Eq.(31) has a zero $\lambda_i$, the second-order term in Eq.(36) is composed only of $(\Delta\psi(x))_{ij}$'s for which Eq.(31) holds. This implies that if we consider a first-order change $\Delta_n\psi(x) = \delta_n\psi(x)$ in Eq.(36) (in the sense of Eq.(31)), its components $(\delta\psi(x))_{ij}$ not vanishing on the right of Eq.(36) because of a zero $\lambda_i$ will represent "directions" that are in the domain Eq.(1). Consequently, the number of those with a negative $\lambda_i$ will give the number of independent directions along which $A[\psi,\psi^*]$ decreases. We can conclude that the index of a given stationary point of $\int \psi^*(x)\hat{A}\psi(x)dx$ under Eq.(1), with an eigenvalue $A_k$, is given by the number of (linearly independent) lower-lying eigenstates of $\hat{A}$, with $A_i < A_k$. Further, as we have seen, all eigenfunctions of Eq.(23) belonging to the zero eigenvalue are eigenfunctions (up to a real multiplier) of the given $\hat{A}$ that correspond to the same eigenvalue $A_k$ as the considered $\psi(x)$; therefore, along a $(\delta\psi(x))_{ij}$ with $\lambda_i = 0$, $A[\psi,\psi^*]$ does not change, that is, does not decrease.

It is worth examining what could be concluded on the considered problem without the second derivative test based on the concept of constrained derivatives. The well-known sufficient condition for a local minimum subject to some constraint $C[\rho]=C$ is that the eigenvalues of

$$\int\left(\frac{\delta^2 A[\rho]}{\delta\rho(x)\delta\rho(x')}-\mu\frac{\delta^2 C[\rho]}{\delta\rho(x)\delta\rho(x')}\right)\Delta\rho(x')dx' = \lambda\Delta\rho(x) \tag{37}$$

be strongly positive, with $\mu = \frac{\delta A[\rho]}{\delta\rho(x'')}\bigg/\frac{\delta C[\rho]}{\delta\rho(x'')}$ being the Lagrange multiplier corresponding to the constraint. However, this is a too strict sufficient condition, being obtained as a consequence of the sufficient condition [5]

$$\iint\left(\frac{\delta^2 A[\rho]}{\delta\rho(x)\delta\rho(x')}-\mu\frac{\delta^2 C[\rho]}{\delta\rho(x)\delta\rho(x')}\right)\Delta_{\bar{C}}\rho(x)\Delta_{\bar{C}}\rho(x')dxdx' \quad \text{be strongly positive}, \tag{38}$$

where $\Delta_{\bar{C}}\rho(x)$ needs not be arbitrary, but satisfies the constraint to first order. That is, there may be a local minimum even in the case of negative eigenvalues among the $\lambda$'s of Eq.(37) (see e.g., [13,14]) – as shown explicitly in the case of Hilbert space functionals by Vogel [14]. Consequently, although the equation $\hat{H}\Delta\psi(x) - E_k\Delta\psi(x) = \lambda\Delta\psi(x)$, arising for Eq.(37) in the case of the problem considered in this study, is just Eq.(24) (obtained with the use of constrained derivatives), the test behind this equation is inconclusive for a spectrum Eq.(26). Eq.(6a) *is what* proves that in the case of both positive and negative eigenvalues appearing in the spectrum of this equation, the examined stationary point is a saddle point. Thus, the extra effort to calculate the constrained second derivatives was not in vain, but to prove that a negative eigenvalue of Eq.(24) signifies a direction along which the value of the considered functional decreases. Note also that in general, Eq.(37) will not lead to the same equation as Eq.(7).

To close this study, we mention that the constrained derivatives corresponding to the constraint of Eq.(1) together with orthogonality to the first *l* eigenstates can be obtained from

$$\psi_{no}[\psi,\psi^*] = \frac{\psi(x)-\psi_1(x)\int\psi_1^*(x')\psi(x')dx'-...-\psi_l(x)\int\psi_l^*(x')\psi(x')dx'}{\sqrt{\int\left|\psi(x'')-\psi_1(x'')\int\psi_1^*(x')\psi(x')dx'-...-\psi_l(x'')\int\psi_l^*(x')\psi(x')dx'\right|^2 dx''}}, \tag{39}$$

inserted into Eq.(10) in the place of Eq.(11). Eq.(39) satisfies the normalization and orthogonality constraints for any $\psi(x)$, and gives back $\psi(x)$ itself when that fulfills those constraints. In this case, the analysis presented in this study will find the Morse index of a given eigenstate shifted by -(*l-s*) (with *s* being the number of states among the "first" *l* eigenstates that correspond to the same energy as the examined eigenstate), due to the additional constraints on the variational domain. (Restricting the domain on which stationarity is considered by constraints reduces the number of independent variational directions, limiting the value of the Morse indices of the stationary points.)

**Acknowledgments:** The author thanks István Mayer, Balázs Pintér, and Péter G. Szalay for valuable discussions. This work was supported by the U.S. Department of Energy TMS program under Grant No. DE-SC0002139.

**Appendix: General method to determine the index of a constrained stationary point**

Inspired by the fact that the eigenvalues of Eq.(23) have proved to be the proper basis for the determination of the index of a stationary point in the case of functionals $A[\psi,\psi^*]=\int \psi^*(x)\hat{A}\psi(x)dx$, in this Appendix, we will establish this result for a general situation, by finding a proper choice to fix the ambiguity in the definition of constrained derivatives.

As has been shown in [11], with the help of constrained derivatives, a similar necessary and sufficient condition, Eqs.(6), can be established for a local minimum under constraint as in the unconstrained case, Eqs.(5). The constrained second derivative entering Eqs.(6) is defined as the unconstrained second derivative of the functional $A[\rho_C[\rho]]$,

$$\frac{\delta^2 A[\rho]}{\delta_C \rho(x) \delta_C \rho(x')} := \frac{\delta^2 A[\rho_C[\rho]]}{\delta \rho(x) \delta \rho(x')}\bigg|_{\rho=\rho_C}, \quad (A1)$$

where the functional $\rho_C[\rho]$ has the following properties: (i) $\rho_C[\rho]$ maps any $\rho(x)$ onto a $\rho_C(x)$, which satisfies the constraint, and (ii) $\rho_C[\rho]$ becomes an identity for $\rho_C(x)$'s, i.e., $\rho_C[\tilde{\rho}_C(x')]=\tilde{\rho}_C(x)$. This definition implies an ambiguity in constrained derivatives, appearing in the form

$$\frac{\delta A[\rho]}{\delta_C \rho(x)} = \frac{\delta A[\rho]}{\delta \rho(x)} - \frac{\delta C[\rho]}{\delta \rho(x)} \int \left( u(x') \bigg/ \frac{\delta C[\rho]}{\delta \rho(x')} \right) \frac{\delta A[\rho]}{\delta \rho(x')} dx' \quad (A2)$$

for the first constrained derivative, where $u(x)$ is an arbitrary function that integrates to 1. It has also been shown that this ambiguity can be embraced by a form

$$\rho_C[\rho] = f^{-1}\left(f(\rho(x)) - u(x)\left(\int f(\rho(x'))dx' - C\right)\right) \quad (A3)$$

for $\rho_C[\rho]$, if the constraint has the form $\int f(\rho(x))dx = C$, with an invertible function $f$. For Eqs.(6), any choice of $\rho_C[\rho]$ can be taken. It is a question, however, whether different choices of $\rho_C[\rho]$ will give a second derivative test of the same value. In the following, this consideration will be applied to choose a particular $u(x)$.

Eqs.(6), more precisely, the necessary condition

$$\iint \frac{\delta^2 A[\rho]}{\delta_C \rho(x)\delta_C \rho(x')} \Delta\rho(x)\Delta\rho(x')dxdx' \geq 0 \quad \text{for all } \Delta\rho(x) \quad (A4)$$

and the sufficient condition

$$\iint \frac{\delta^2 A[\rho]}{\delta_C \rho(x)\delta_C \rho(x')} \Delta\rho(x)\Delta\rho(x')dxdx' > p\|\Delta\rho\| \quad \text{for all nonzero } \Delta\rho(x) \quad (A5)$$

for a local minimum under constraint can be derived from the necessary condition [5]

$$\iint \left(\frac{\delta^2 A[\rho]}{\delta\rho(x)\delta\rho(x')} - \mu\frac{\delta^2 C[\rho]}{\delta\rho(x)\delta\rho(x')}\right)\Delta_{\bar{C}}\rho(x)\Delta_{\bar{C}}\rho(x')dxdx' \geq 0 \quad \text{for all } \Delta_{\bar{C}}\rho(x) \quad (A6)$$

and sufficient condition [5]

$$\iint \left(\frac{\delta^2 A[\rho]}{\delta\rho(x)\delta\rho(x')} - \mu\frac{\delta^2 C[\rho]}{\delta\rho(x)\delta\rho(x')}\right)\Delta_{\bar{C}}\rho(x)\Delta_{\bar{C}}\rho(x')dxdx' > p\|\Delta_{\bar{C}}\rho\| \quad \text{for all nonzero } \Delta_{\bar{C}}\rho(x) \quad (A7)$$

respectively. In Eqs.(A5) and (A7), $p$ is some positive number (due to the strong positivity requirement), while in Eqs.(A6) and (A7), $\Delta_{\bar{C}}\rho(x)$ signifies increments that satisfy the constraint to first order, i.e.,

$$\int \frac{\delta C[\rho]}{\delta\rho(x)}\Delta_{\bar{C}}\rho(x) = 0 \; . \quad (A8)$$

The essence of the derivation of Eqs.(A4) and (A5) was to prove

$$\iint \left(\frac{\delta^2 A[\rho]}{\delta\rho(x)\delta\rho(x')} - \mu\frac{\delta^2 C[\rho]}{\delta\rho(x)\delta\rho(x')}\right)\Delta_{\bar{C}}\rho(x)\Delta_{\bar{C}}\rho(x')dxdx' = \iint \frac{\delta^2 A[\rho]}{\delta_C\rho(x)\delta_C\rho(x')}\Delta\rho(x)\Delta\rho(x')dxdx', \quad (A9)$$

with

$$\Delta_{\bar{C}}\rho(x) = \int \frac{\delta\rho(x)}{\delta_C\rho(x')}\Delta\rho(x')dx' \; . \quad (A10)$$

Then, the equivalence of Eq.(A4) with Eq.(A6) is obvious, while Eq.(A5) trivially implies Eq.(A7), since (i) Eq.(A5) is required for $\Delta\rho(x) = \Delta_{\bar{C}}\rho(x)$, too, (ii)

$$\frac{\delta^2 A[\rho]}{\delta_C \rho(x) \delta_C \rho(x')} = \iint \frac{\delta \rho(x'')}{\delta_C \rho(x)} \left( \frac{\delta^2 A[\rho]}{\delta \rho(x'') \delta \rho(x''')} - \mu \frac{\delta^2 C[\rho]}{\delta \rho(x'') \delta \rho(x''')} \right) \frac{\delta \rho(x''')}{\delta_C \rho(x')} dx'' dx''' \ , \tag{A11}$$

and (iii) Eq.(A10) is an identity for $\Delta \rho(x) = \Delta_{\bar{C}} \rho(x)$'s. Since Eq.(A4) excludes the possibility of a negative second differential at a local minimum, the previous argument already implies that Eq.(A5) is a strong enough sufficient condition for a local minimum, leaving only the situation of the second differential equaling zero as an inconclusive case (similar to the unconstrained version, Eqs.(5)). To establish the equivalence of Eq.(A5) with Eq.(A7) (i.e., Eq.(A7)=>Eq.(A5)), we need to show that when Eq.(A5) is inconclusive, Eq.(A7) is inconclusive, too. Eq.(A5) can be inconclusive in two cases – when (i) for some $\Delta \rho(x)$, the second differential in Eq.(A5) vanishes, or (ii) for some sequence of $\Delta \rho(x)$'s, that second differential tends to zero. The first case implies on the basis of Eq.(A9) that there exists a $\Delta_{\bar{C}} \rho(x)$ (obtained from the given $\Delta \rho(x)$ via Eq.(A10)) for which the second differential in Eq.(A7) vanishes; consequently, Eq.(A7) is also inconclusive. In case (ii), the given sequence of $\Delta \rho(x)$'s generates a corresponding sequence of $\Delta_{\bar{C}} \rho(x)$'s (via Eq.(A10)) for which the second differential in Eq.(A7) tends to zero, implying an inconclusiveness. Thus, Eqs.(A5) and (A7) are equivalent.

To find the proper choice of $u(x)$ to determine the indices of stationary points, the following, basic property of constrained derivatives will be utilized:

$$\int \left( u(x) \bigg/ \frac{\delta C[\rho]}{\delta \rho(x)} \right) \frac{\delta A[\rho]}{\delta_C \rho(x)} dx = 0 \ , \tag{A12}$$

which follows from Eq.(A2) directly. Applying Eq.(A12) to the first constrained derivative itself, we have

$$\int \left( u(x) \bigg/ \frac{\delta C[\rho]}{\delta \rho(x)} \right) \frac{\delta^2 A[\rho]}{\delta_C \rho(x) \delta_C \rho(x')} dx = 0 \ . \tag{A13}$$

Now, use of this equation can be made on the eigenvalue equation Eq.(7) to obtain

$$\lambda \int \left( u(x) \bigg/ \frac{\delta C[\rho]}{\delta \rho(x)} \right) \Delta \rho(x) dx = 0 \tag{A14}$$

as a property of its eigenfunctions and corresponding eigenvalues. Eq.(A14) shows us that the choice

$$u(x) = \left( \frac{\delta C[\rho]}{\delta \rho(x)} \right)^2 \bigg/ \int \left( \frac{\delta C[\rho]}{\delta \rho(x')} \right)^2 dx' \tag{A15}$$

is special, since with Eq.(A15), Eq.(A14) gives that any eigenfunction of Eq.(7) corresponding to a nonzero $\lambda$ will satisfy Eq.(A8) (i.e. will be a $\Delta_{\bar{C}}\rho(x)$). This is important because with the use of the expansion of a general $\Delta\rho(x)$ in terms of orthogonal eigenfunctions of Eq.(7), a general change of $A[\rho]$ over the domain of $C[\rho]=C$ emerges as

$$A[\rho+\Delta_C\rho]-A[\rho] = \sum_{i,j}\lambda_i c_{ij}^2 \int (\Delta\rho(x))_{ij}^2 dx + higher-order\,terms \ . \qquad (A16)$$

With Eq.(A15), the second-order term on the right of Eq.(A16) is composed only of terms with $(\Delta\rho(x))_{ij}$ satisfying Eq.(A8). Consequently, the set of $\lambda$'s obtained with the use of Eq.(A15) is a proper basis for the determination of the Morse index under constraint: the number of negative $\lambda$'s will give the index.